\documentclass[lettersize,journal]{IEEEtran}
\usepackage{amsmath,amsfonts, amssymb}
\usepackage{algorithmic}
\usepackage{array}
\usepackage[caption=false,font=normalsize,labelfont=sf,textfont=sf]{subfig}
\usepackage{textcomp}
\usepackage{stfloats}
\usepackage{url}
\usepackage{verbatim}
\usepackage{graphicx}
\usepackage{color}
\usepackage{hyperref}
\usepackage{cleveref}
\usepackage{booktabs}
\usepackage{soul}
\usepackage{bm}

\usepackage[numbers,sort&compress]{natbib}

\usepackage{tikz}
\usepackage{textcomp}

\newcommand{\bl}[1]{{\color{black} #1}}

\newcommand{\black}[1]{{\color{black} #1}}

\newcommand{\blu}[1]{{\color{black} #1}}

\def\BibTeX{{\rm B\kern-.05em{\sc i\kern-.025em b}\kern-.08em
    T\kern-.1667em\lower.7ex\hbox{E}\kern-.125emX}}
\usepackage{balance}
\begin{document}
\title{Overview of Automatic Speech Analysis and Technologies for Neurodegenerative Disorders: Diagnosis and Assistive Applications}

\author{Shakeel A.~Sheikh, Md. Sahidullah,~\IEEEmembership{Member,~IEEE}  and Ina Kodrasi,~\IEEEmembership{Senior Member,~IEEE} 

\thanks{Shakeel A.~Sheikh and Ina Kodrasi are with Idiap Research Institute, Switzerland (e-mail: \{shakeel.sheikh, ina.kodrasi\}@idiap.ch).}

\thanks{Md Sahidullah is with TCG CREST, Kolkata, India (e-mail: md.sahidullah@tcgcrest.org).}

\thanks{This work was supported by the Swiss National Science Foundation project no CRSII5\_202228 on ``Characterisation of motor speech disorders and processes''.}

}

\markboth{IEEE Journal of Selected Topics in Signal Processing, ~Vol.~XX, No.~XX, December~2024}%
{How to Use the IEEEtran \LaTeX \ Templates}

\maketitle

\begin{abstract}
\black{
Advancements in spoken language technologies for neurodegenerative speech disorders are crucial for meeting both clinical and technological needs. This overview paper is vital for advancing the field, as it presents a comprehensive review of state-of-the-art methods in pathological speech detection, automatic speech recognition, pathological speech intelligibility enhancement, intelligibility and severity assessment, and data augmentation approaches for pathological speech. It also highlights key challenges, such as ensuring robustness, privacy, and interpretability. The paper concludes by exploring promising future directions, including the adoption of multimodal approaches and the integration of large language models to further advance speech technologies for neurodegenerative speech disorders. 

}
\end{abstract}


\section{Introduction}
\label{intro}

\IEEEPARstart{S}{peech} production is a complex mechanism that involves cognitive planning, coordinated muscle activity, and sound creation~\cite{simonyan_neuro_2016}.
The process starts in the brain with the conceptualization of a message, followed by the organization of phonetic and prosodic plans, such as rhythm and style.
The motor cortex then orchestrates the activation of approximately $100$ muscles, allowing articulatory organs such as the tongue, lips, and jaw to shape the vocal tract and produce specific sounds.
The initial sound is generated in the larynx, where the air from the lungs causes the vocal folds to vibrate. These phonatory structures adjust voice quality and prosody, while the articulatory organs further refine the sound by altering the shape of the vocal tract. Finally, the resulting speech is emitted through the oral and nasal cavities.
Given the intricate coordination required for this process, any disruption of these finely tuned mechanisms can severely alter communication and result in pathological speech.

In this work, we define pathological speech as speech that deviates from neurotypical patterns due to underlying impairments. These deviations can manifest along multiple dimensions, including voice, articulation, prosody, and language. Voice impairments involve abnormalities in vocal fold vibration or in breath control, leading to hoarseness, breathiness, or a strained voice~\cite{sewall_2024, isshiki_differential_1969}. Articulation impairments involve abnormalities in the coordination or movement of the various articulators, leading to slurred, imprecise, or segmented speech~\cite{duffy_motor, aos_overview}. Furthermore, prosodic impairments involve abnormalities in the rhythm, stress, or intonation of speech, leading to speech that may sound flat or monotonic~\cite{troger2024automatic}. Finally, language impairments involve abnormalities in the formulation or comprehension of linguistic content, leading to difficulties with word retrieval, sentence construction, or understanding spoken or written language~\cite{aphasia}. While pathological speech can arise from a wide range of neurological, structural, or functional impairments, our objective is to provide an overview of automated methods and speech-based technologies targeting pathological speech arising due to neurodegenerative disorders.

Neurodegenerative disorders such as Parkinson's disease (PD), Amyotrophic Lateral Sclerosis~(ALS), or Alzheimer's disease are leading causes of voice, articulation, prosody, and language disruptions~\cite{damico2010handbook,slp_book,duffy_motor}.
These disorders impair the brain regions and motor systems responsible for initiating, planning, and controlling the movements needed for speech production, resulting in a variety of speech disorders such as dysarthria, aphasia, apraxia of speech, or dysphonia~\cite{aos_overview,sewall_2024, ars,duffy_motor,aphasia,stuttering, sataloff_clinical_2017}.
\bl{Dysarthria and apraxia of speech, commonly seen in PD and ALS, are primarily characterized by articulatory and prosodic deficiencies}~\cite{moro-velazquez_forced_2019}, vowel distortions, reduced loudness variation, hypernasality, or syllabification~\cite{aos_overview,duffy_motor}.
Dysphonia, also frequent in PD and ALS, is marked by abnormal voice quality such as hoarseness and breathiness~\cite{sewall_2024, isshiki_differential_1969}.
In contrast, aphasia typically presents as difficulties with word-finding and sentence construction, and is most commonly associated with Alzheimer's disease or other forms of dementia~\cite{aphasia}.

\blu{
As the population grows and ages, the prevalence of neurological disorders, and consequently of various speech disorders, rapidly increases.} In $2019$, the World Health Organization estimated that over $8.5$ million people worldwide were living with PD~\cite{who}, up from $6.1$ million in $2016$~\cite{pd_incidence} and $2.5$ million in $1990$~\cite{pd_incidence}. 
By $2040$, this figure is projected to surpass $17$ million~\cite{pd_incidence}.
Similarly, dementia affected more than $46$ million people globally in $2015$ and this figure is expected to rise to $131.5$ million by $2050$~\cite{dementia_increase}. 
The prevalence of ALS is also growing significantly, with cases anticipated to increase by nearly 70\% between $2015$ and $2040$~\cite{als_increase}. 
This increasing prevalence of neurological disorders, and consequently of the associated speech disorders, underscores the need to prioritize speech disorders both in the context of clinical practice as well as in the context of speech-based technologies.

\blu{Accurately diagnosing the presence of speech disorders in clinical practice (i.e., distinguishing between neurotypical and impaired speech) is crucial, as the presence of such disorders may serve as an early indicator of neurodegenerative conditions~\cite{Rusz_JASA_2013,Rong_2015,tracy_bi_2020}.}
Further, an accurate differential diagnosis (e.g., discriminating between dysarthria and apraxia of speech) can provide important clues about the underlying neuropathology~\cite{Duffy_2008,Duffy_book_2000}. Monitoring speech characteristics such as severity and intelligibility after diagnosis is also essential for tracking disease progression and evaluating the effectiveness of speech therapy interventions over time~\cite{damico2010handbook, kempster2009consensus, sataloff_clinical_2017, troger2024automatic}.

Speech assessment in clinical practice relies on established perceptual evaluation scales that serve as gold standards for diagnosing and characterizing various aspects of speech impairments. For example, the GRBAS scale~\cite{omori2011diagnosis, saenz-lechon_automatic_2006} evaluates voice quality through parameters like grade, roughness, breathiness, asthenia, and strain. The Consensus Auditory Perceptual Evaluation-Voice (CAPE-V)~\cite{kempster2009consensus} offers a similar perceptual framework, excluding asthenia.
In addition to these general assessment scales, specific scales are employed for specific conditions. 
For instance, the Unified Parkinson's Disease Rating Scale (UPDRS)~\cite{updrs} includes components for speech and motor function assessment in PD, while the Bogenhausen Dysarthria Scales (BoDyS)~\cite{bodys} focus on the severity and profile of dysarthric impairments. Intelligibility, a key outcome measure in many disorders, is often assessed using tools such as the Assessment of Intelligibility of Dysarthric Speech~\cite{yorkston_assessment_1981}, which standardizes both single-word and sentence intelligibility evaluations.

\begin{figure*}
    \centering
    \includegraphics[scale=1]{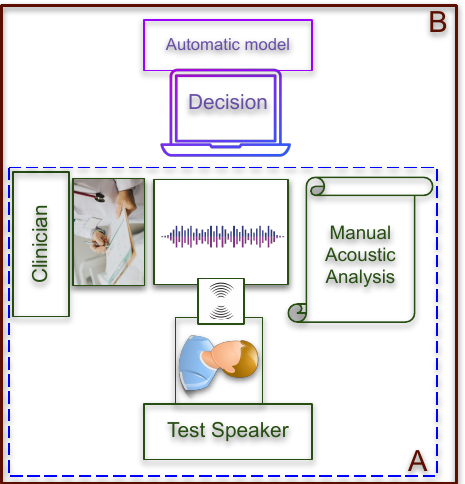}
    \caption{\bl{Traditional auditory-perceptual assessment in clinical practice (bounded by the dashed box A) and automatic pathological speech analysis system (bounded by the solid box B). The  clinician listens to the (potential) patient and assesses by ear the various characteristics of the speech. The automatic model is trained to detect and analyze speech impairments. Clinicians may use the insights provided by the automatic model to organize therapeutic sessions accordingly. Additionally, they may perform manual acoustic analysis to gain further insights into the patient's speech patterns and impairments, providing complementary information to the automatic model's decision.}}
    \label{fig:manual}
\end{figure*}

Traditionally, clinicians conduct these evaluations through costly and time-consuming auditory-perceptual assessments~\cite{sataloff_clinical_2017, kempster2009consensus}, as illustrated in Fig.\ref{fig:manual}. 
\blu{Diagnosing speech disorders and distinguishing between various conditions can be particularly challenging, even for experienced clinicians. This difficulty arises from the subtle nature of clinical-perceptual characteristics which are often hard to detect by ear, especially in cases of mild impairments. The overlapping characteristics of certain speech disorders, such as dysarthria and apraxia of speech, further complicate the process. Consequently, inter-rater agreement for (differential) diagnosis of speech disorders among clinicians can be low~\cite{Bunton_JSLHR_2008,Fonville_JN_2008}.}

These clinical challenges highlight the growing need for complementary, technology-driven approaches to support diagnosis, monitoring, and intervention. In parallel, the increasing prevalence of speech disorders poses significant accessibility challenges in patients' everyday interactions with speech-based technologies. For example, individuals with dysarthria or apraxia of speech often experience difficulty using mainstream virtual assistants such as Cortana, Alexa, and Siri~\cite{de2019impact}.
As speech disorders become more prevalent with neurodegenerative conditions, it is critical to prioritize both the diagnosis and treatment of these disorders in clinical practice, while also ensuring that patients with such impairments have equitable access to speech-based technologies. Addressing these barriers could lessen the burden on the health care system and significantly improve the patients' quality of life and their ability to engage with everyday digital tools.

\begin{figure*}[ht]
    \centering
    
    \includegraphics[scale=0.8]{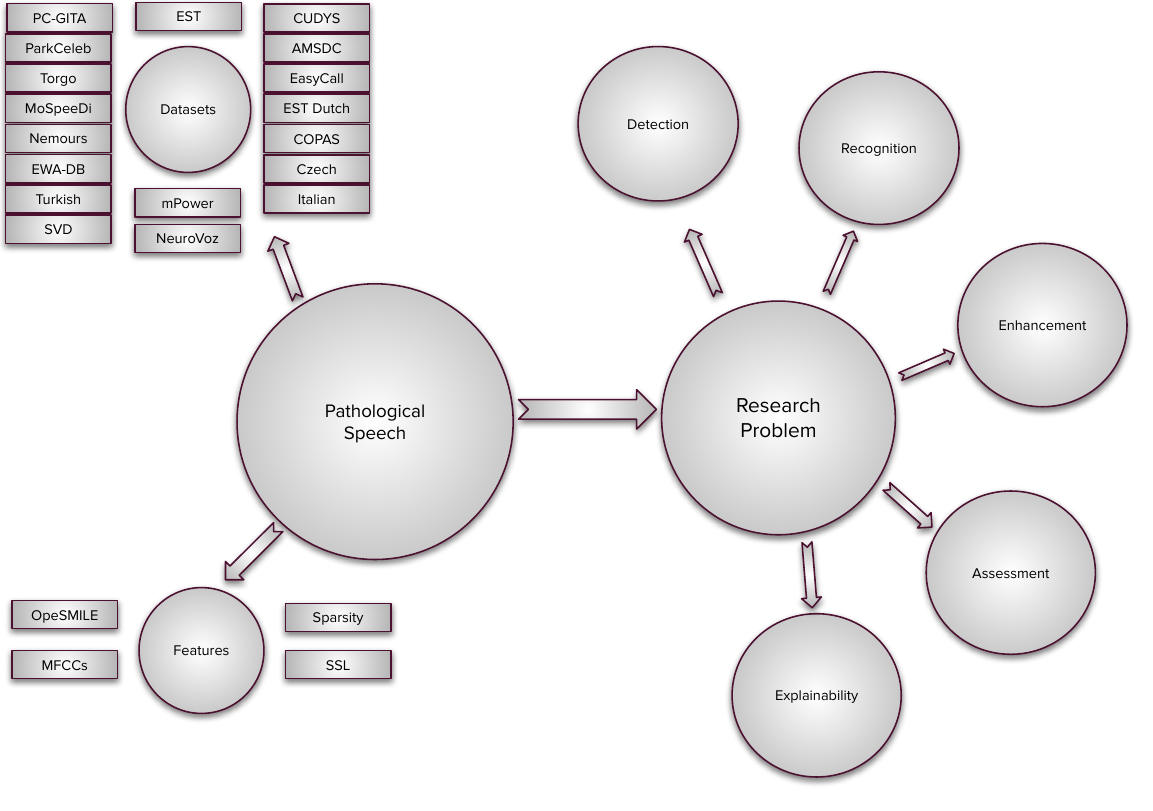}
    \caption{Overview of key components discussed in this manuscript, including datasets, features, and research directions for pathological speech.}
    \label{fig:ai4pd}
\end{figure*}

Aiming to assist the clinical diagnosis and treatment
of patients suffering from neurodegenerative disorders, there has 
been a growing interest in the research community to develop automated and objective methods for pathological speech analysis. A schematic illustration of such analysis is depicted in Fig. \ref{fig:manual}. These advanced technologies are designed to minimize bias, enhance diagnostic accuracy, and streamline the assessment process, ensuring greater efficiency and consistency. 
Clinicians can then use insights provided by the automatic models to organize therapeutic sessions accordingly, ensuring that the treatment addresses the specific impairments identified. Further, clinicians may perform additional manual acoustic analysis to gain further insights into the patient's speech patterns and impairments, providing complementary information to the automatic model's decision for a comprehensive therapeutic approach.
Besides the clinical domain, efforts have been directed towards developing various speech-based technological applications aimed at pathological speakers, such as automatic speech recognition systems (ASRs) \cite{liu_recent_2021}, speech synthesis systems \cite{kain04_ssw, soleymanpour_synthesizing_2022, veaux_using_2012, dhanalakshmi_speech-input_2018} or intelligibility enhancement solutions \cite{yang_improving_2020, halpern_objective_2021, chandrashekar_spectro-temporal_2020, tripathi_improved_2020}.
To our knowledge, there is currently no comprehensive survey paper that discusses the research directions and challenges of this area both from a clinical and a technological perspective.
\blu{
Although, there have been related works, such as \cite{gomez2019design, moro-velazquez_advances_2021, ding2024speech}, they are primarily focused on specific characteristics, applications, or disorders, limiting their scope.}
For instance, \cite{rowe2023quantifying} provides an overview of acoustic-articulatory characteristics in neurodegenerative disorders. Other studies, such as \cite{van2024innovative, cordella2024introduction, brahmi2024exploring, gupta2024voice, rehman2024voice, amato2023machine}, mainly focus their review on the discrimination between neurotypical and impaired speech. \citet{gupta2016pathological} discussed some of the wider challenges faced in the pathological speech domain. However, this work addresses only a limited set of challenges and is now somewhat outdated given the rapid advancements in the field over the past decade. In contrast, as depicted in Fig. \ref{fig:ai4pd}, our work aims to fill this gap by providing an extensive review of the field encompassing pathological speech from various clinical and technological perspectives, such as automatic discrimination between neurotypical speech and speech disorders, ASR systems for pathological target speakers, enhancement systems aiming to enhance the intelligibility of pathological speech, severity and intelligibility assessment, and data augmentation approaches.

The remainder of the paper is organized as follows. \Cref{datasets} describes various pathological datasets employed in the literature. \Cref{feats} provides a high level overview of the various speech representations used in pathological speech analysis. \Cref{detection} describes the different approaches to automatic pathological speech detection. \Cref{pasr} examines pathological speech in the context of ASR systems. \Cref{enhacement} discusses speech enhancement techniques aimed at improving pathological speech intelligibility.~\Cref{intelli_assessment} describes approaches for automatically estimating the intelligibility and severity of pathological speakers.~\Cref{dataaug} summarizes the various data augmentation methods used in automatic pathological speech systems. 
Finally, \Cref{challenges} presents challenges and promising future research directions in the field.
In summary, our contributions are the following:
\begin{enumerate}
    \item \black{We present the first comprehensive survey of automatic approaches for pathological speech from a clinical and technological perspectives ranging from detection, recognition, enhancement, and assessment}.
    \item We highlight current limitations in the field and propose several promising future directions for research in pathological speech processing. 
\end{enumerate}

\section{Pathological Speech Datasets}
\label{datasets}

Datasets for pathological speech research are scarce due to several inherent challenges in data collection. One of the primary difficulties is the sensitive nature of the population involved. Recruiting participants suffering from neurological disorders requires careful consideration of ethical and privacy concerns, as well as navigating the potential stigma associated with such disorders. Additionally, the physical and cognitive challenges faced by these individuals can complicate the process of obtaining high-quality speech recordings.
Another key challenge is the variability in pathological speech patterns. Speech impairments can manifest in numerous ways and may vary widely across individuals, even within the same diagnostic category. This variability makes it difficult to create standardized protocols for data collection that ensure consistency and relevance across samples. Moreover, speech impairments may fluctuate over time, further complicating the process of capturing representative speech samples.
When collecting datasets for pathological speech research, it is essential to prioritize inclusivity, i.e., ensuring a diverse representation of different speech disorders, age groups, and demographics. Additionally, data collection protocols should consider the comfort and cooperation of participants. Ethical considerations must also be at the forefront, ensuring informed consent and the protection of participant privacy. Lastly, it is important to design flexible and scalable collection methods that can capture a range of speech characteristics while maintaining consistent quality across diverse individuals and conditions.
In the remainder of this section, we briefly review  pathological datasets\footnote{Due to restricted access to many datasets, we are unable to provide complete metadata, including details like the number of male and female speakers} commonly used in the literature. The summary of these datasets and their characteristics is presented in Table \ref{tab:pathological_speech_databases}.

\begin{itemize}

    \item \emph{TORGO ~\cite{rudzicz_torgo_2012}.} \enspace The TORGO dataset contains English \blu{(spontaneous and read)}  speech recordings from control speakers and patients and the corresponding three-dimensional \blu{electromagnetic articulography (EMA)}. The patients suffer from ALS or Cerebral Palsy (CP). The dataset consists of recordings from $7$ ($3$ female, $4$ male) control speakers and $8$ ($3$ female, $5$ male) patients.  

\begin{table*}
\centering
\caption{\bl{Overview of pathological speech datasets (PD: Parkinson's disease, CVA: Cerebrovascular accident, TBI: Traumatic brain injuries, ALS: Amyotrophic lateral sclerosis, Huntington’s disease, AD: Alzheimer’s Disease, MCI: Mild Cognitive Impairment, EMA: Electromagnetic Articulography, EGG: Electroglottography, L: Longitudinal data). Public datasets are openly available on the web and can be downloaded without the need for explicit approval or signing an agreement. Accessible datasets require an application process or signing an agreement, but can be obtained by the research community under defined conditions. Non-accessible datasets are private and not available to the wider research community.}}
\blu{
\begin{tabular}{lrrllll}
\toprule
\textbf{Database} & \multicolumn{2}{c}{\textbf{Number of Speakers}} & \textbf{Type of Impairment} & \textbf{Language} & \textbf{Modality} &\textbf{Accessibility} \\ \cmidrule(lr){2-3}
                  & \textbf{Control} & \textbf{Pathological} & & & \\ 
                  \midrule
PC-GITA   & 50  & 50  & PD  & Spanish  & Speech   &Accessible\\
TORGO & 07& 08 & ALS/CP & English& Speech + EMA &Public\\
MoSpeeDi &466 &138 &Dysarthira/Apraxia &French &Speech&Not Accessible \\
Nemours & -&11 &Dysarthira& English&Speech&Not Accessible \\
CUDYS & 05 & 11 & Spino-cerebellar ataxia&Cantonese &Speech&Not Accessible \\
AMSDC &62 &37 &CVA, PD &English & Speech&Not Accessible\\
Dutch EST & -&16 &Dysarthria, TBI
  &Dutch &Speech &Not Accessible \\
EasyCall &24 &31 &PD, HD, ALS &Italian &Speech&Public \\
Saarbrucken Voice Database& 869&1,356 &Pathological & German & Speech + EGG & Public\\
Turkish Parkinson Dataset& 20 & 20 & PD & Turkish & Speech & Not Accessible\\
Czech Parkinson's Dataset& 22 & 61 & PD & Czech & Speech & Public \\
Italian Parkinson's Database &28 &37 & PD &Italian &Speech & Public \\
NeuroVoz& 58 & 54 & PD & Spanish & Speech & Public \\
COPAS&197&122&Dysarthira and others&Dutch&Speech&Accessible \\
EWA-DB & 896 & 226 & PD, AD, MCI & Slovak & Speech & Accessible \\
ParkCeleb&40&40&PD&English&Speech + Video + L&Accessible\\
mPower & 5,581& 1,087& PD & English & Multimodal & Accessible\\
\bottomrule
\end{tabular}
}
\label{tab:pathological_speech_databases}
\end{table*}

\item  \emph{PC-GITA~\cite{orozco-arroyave-etal-2014-new}.} \enspace The PC-GITA dataset contains Spanish \blu{(spontaneous and read)} speech recordings from $50$ control speakers and $50$ patients suffering from PD. The two groups of speakers are age- and sex-matched, with $25$ male and $25$ female speakers in each group.
 \item \emph{MoSpeeDi~\cite{mospeedi}.} \enspace The MoSpeeDi dataset contains French \blu{(spontaneous and read)} speech recordings from $466$ control speakers and $138$ patients suffering from various types of motor speech disorders such as dysarthria or apraxia of speech. While subgroups of age- and sex-matched controls and patients can be found within the database, the overall dataset is not age- and sex-matched.

\item \emph{Nemours~\cite{menendez-pidal_nemours_1996}.} \enspace
The Nemours dataset is an English dataset containing \blu{(read)} speech recordings from $11$ male speakers with varying degrees of dysarthria severity. 

\item \emph{CUDYS~\cite{wong_development_2015}.} \enspace
The Cantonese Dysarthric Speech Corpus contains Cantonese \blu{(read)} speech recordings from $5$ control speakers and $11$ patients suffering from cerebellar degeneration.
The control group consists of $2$ female and $3$ male speakers, whereas the dysarthric group consists of $5$ female and $6$ male speakers. 

\item \emph{AMSDC~\cite{laures-gore_atlanta_2016}.} \enspace
The Atlanta Motor Speech Disorders Corpus is an English dataset containing \blu{(spontaneous and read)} speech recordings from $99$ ($62$ male, $37$ female) patients diagnosed with various disorders such as PD, ALS, or dementia. 

\item \emph{EST~\cite{yilmaz_dutch_2016}.} \enspace
The EST dataset is a Dutch dataset containing \blu{(read)} speech recordings from $16$ male dysarthric patients due to PD, traumatic brain injuries, or cerebrovascular accident.
. 

\item \emph{EasyCall~\cite{turrisi_easycall_2021}.} \enspace
The EasyCall dataset is an Italian dataset containing \blu{(read)} speech recordings  from $24$ controls and $31$ patients diagnosed with PD, Huntington's disease, ALS, and peripheral neuropathy.
The control group consists of $10$ female and $14$ male speakers, whereas the patient group consists of $11$ female and $20$ male speakers.

\item \emph{COPAS \cite{van2009dutch}.} \enspace  The Corpus of Pathological and Normal Speech dataset is a Dutch dataset containing (spontaneous and read) speech recordings from $197$ pathological speakers and $122$ control speakers. The database comprises $8$ distinct pathological categories such as dysarthria, hearing impairment, cleft, etc. 

\item \emph{ParkCeleb \cite{favaro2024unveiling}.} \enspace
The previously reviewed datasets are not longitudinal and do not allow tracking the progression of the speech disorder within the same patient along time. To address this gap, the English ParkCeleb dataset was recently introduced in \cite{favaro2024unveiling}.
\blu{This dataset contains (spontaneous) audio-visual recordings (such as studio interviews or press conferences)} from $40$ ($2$ female, $38$ male) control speakers and $40$ ($2$ female, $38$ male) patients suffering from PD.

\bl{
\item \emph{Italian Parkinson's Database \cite{italian_database}}.
The dataset contains \blu{Italian (read) speech} recordings from a total of $65$ speakers, including $28$ patients with PD and $37$ control speakers. Sex distribution is non-uniform across groups, with the PD group consisting of $19$ male and $9$ female speakers and the control group consisting of $23$ male and $14$ female speakers.

%Public
\item \emph{NeuroVoz \cite{mendes2024neurovoz}}. 
The NeuroVoz dataset contains (\blu{spontaneous, read, and listen and repeat}) speech recordings from $112$ Castilian Spanish speakers, including $54$ patients diagnosed with PD and $58$ control speakers. The control group consists of $28$ male speakers, $26$ female speakers, and $1$ speaker whose sex information is not provided, whereas the patient group consists of $33$ male and $20$ female speakers.

\blu{
\item \emph{Saarbrücken Voice Database \cite{putzer_saarbrucken_2021}\footnote{\url{https://stimmdb.coli.uni-saarland.de/} (accessed July 07, 2025)}}. 
The Saarbr\"{u}cken Voice Database contains (read) speech recordings and electroglottography data from $2,255$ German speakers, including $1,356$ patients and $869$ control speakers~\cite{martinez2012voice}. Patients are diagnosed with various voice disorders.
The control group consists of $433$ male and $436$ female speakers, whereas the patient group consists of $629$ male and $727$ female speakers.
}

\item \emph{Turkish Parkinson Speech Dataset \cite{sakar2013collection}}. This dataset contains \blu{(read)} speech recordings from $40$ Turkish speakers, including $20$ controls and $20$ patients with PD. The PD group contains $6$ female and $14$ male speakers, whereas the control group contains $10$ female and $10$ male speakers.

\item \emph{Czech Parkinson's Dataset \cite{czech}}. 
This dataset contains \blu{(read)} speech recordings from $83$ Czech speakers, including $22$ control speakers and $61$ patients diagnosed with PD or atypical parkinsonian syndromes.
The patient group consists of $30$ female and $31$ male speakers, whereas the control group consists of $11$ female and $11$ male speakers. 
}

\item \emph{EWA-DB \cite{rusko2024slovak}}.
The EWA-DB dataset consists of \blu{(spontaneous and read)} speech recordings of $1,122$ Slovak speakers, including $896$ control speakers and $226$ patients diagnosed with PD, mild cognitive impairment, or Alzheimer's disease. 
The patient group consists of $121$ male and $105$ female speakers, whereas the control group consists of $248$ male and $648$ female speakers. 
\textcolor{black}{
\item \emph{mPower Parkinson's Dataset}~\cite{bot2016mpower}. The mPower study collected multimodal smartphone sensor data from $6805$ participants, including $1,087$ patients diagnosed with PD and $5,581$ control speakers.  Patient and control status were self-reported. The dataset was collected through the mPower mobile application and includes four distinct sensor-based assessment modalities, i.e., spatial memory evaluation, gait analysis through walking tasks, manual dexterity measurement via finger tapping, and vocal function assessment using sustained phonation recordings.
}
\end{itemize}

\blu{As presented in Table \ref{tab:pathological_speech_databases}, a considerable number of the datasets used in the literature are private and not openly accessible. This limits their availability for broader research and replication efforts. Although some datasets such as TORGO are publicly available, they suffer from detrimental recording artifacts \cite{janbakhshi_ua}, which compromise their usefulness for studying pathological speech characteristics and for developing reliable automatic detection methods \cite{janbakhshi_ua, amiri_iwanc}. The lack of accessible, high-quality data poses a major challenge to the development and validation of effective algorithms, slowing progress in understanding and addressing pathological speech conditions. Ensuring the availability of clean, reliable, and comprehensive datasets is therefore essential for advancing research and practical applications.

Another critical limitation is the lack of linguistic diversity. Existing datasets are available only in a few languages, which restricts their applicability to non-represented populations and limits the potential for cross-linguistic comparisons. This linguistic bias poses a challenge for developing universal models that can generalize across languages.

In addition to these issues, most datasets are relatively small, often including only a limited number of participants. This scarcity hinders the development of robust and generalizable models and makes it difficult to capture the wide variability in speech disorders across different individuals and conditions. Furthermore, demographic imbalances are common. Many datasets exhibit skewed gender representation, which may introduce biases into trained models. In some cases, age groups or types of pathological conditions are also underrepresented, further limiting generalizability.

These limitations underscore the pressing need for larger, more diverse, and better-balanced pathological speech datasets that reflect the complexity of real-world populations. Addressing these gaps is essential for building reliable and inclusive tools for pathological speech analysis.}

\section{Speech Representations for \\ Automatic Approaches}
\label{feats}

Pathological speech exhibits a range of acoustic anomalies, including deviations in pitch, loudness, vowel space reduction, and articulation~\cite{lansford2014vowel, duffy_motor}.
Additionally, it can lead to asymmetrical tension in the vocal folds, resulting in irregular vibrations and, consequently, an abnormal fundamental frequency. 
Patients can also show inconsistent rhythmic structures in comparison to control groups~\cite{pina2024speech}.
Excessively high or low fundamental frequency, combined with excessive vocal intensity, can exacerbate the severity of pathological voice conditions, producing characteristics such as shrillness, screechiness, hoarseness, or huskiness~\cite{davis1979acoustic}.

To capture such abnormalities as biomarkers for automatic pathological speech processing, researchers have employed various handcrafted acoustic features such as OpenSMILE \cite{janbakhshi_ua}, Mel-frequency cepstral coefficients (MFCCs) \cite{orozco2015voiced}, or spectro-temporal sparsity features \cite{kodrasi_spectro-temporal_2020}. 
Raw speech signals and various time-frequency representations such as the short-time Fourier transform (STFT) have also been directly exploited in combination with deep learning approaches to directly learn pathology-discriminant cues \cite{janbakhshi_experimental_2022,rawspeech1}.
More recently, latent embeddings derived from self-supervised models have been used as more powerful representations of speech patterns and the various impairments \cite{janbakhshi_ua, javanmardi_wav2vec-based_2023, violeta22_interspeech}.
It is important to note that a considerably larger set of features and representations have been employed to characterize disorders across various pathologies than the ones outlined above. Here, we briefly discuss the representations that are most commonly used in the literature. 

\subsection{Handcrafted Acoustic Features}
\paragraph*{OpenSMILE}\black{Among the various features used in pathological speech detection, OpenSMILE features have been widely explored in the literature in conjunction with traditional machine learning algorithms~\cite{eyben2010opensmile, prabhakera_dysarthric_2018, millet_learning_2019, narendra_dysarthric_2019, tripathi_improved_2020, alhinti_recognising_2020, descrimination_ina, narendra_automatic_2021, kodrasi_automatic_2021, tripathi_automatic_2021, janbakhshi_ua, joshy_automated_2022, javanmardi_pre-trained_2024, speech_mode_impact}. The OpenSMILE feature set includes a $6552$-dimensional feature vector that primarily consists of low-level audio features such as CHROMA, CENS, loudness, MFCCs, and other spectral features. In the context of pathological speech, these features can capture the subtle anomalies in speech patterns that are indicative of disorders. OpenSMILE's comprehensive feature set allows for the analysis of prosodic elements like pitch, jitter, shimmer, and formant frequencies, which are crucial for identifying and differentiating various speech pathologies. However, OpenSMILE features are general features that have been used for a variety of speech applications such as emotion recognition~\cite{liu2023paralinguistic} or speech recognition~\cite{toyama2017use} and they are not specifically handcrafted to capture pathological cues.}.

\paragraph*{Spectral and cepstral coefficients} 
Due to their ability to characterize articulation deficiencies, various spectral and cepstral coefficients such as Linear Predictive Coding (LPC), MFCCs, and Perceptual Linear Prediction (PLP) features, have been successfully used for pathological speech detection~\cite{shahamiri_artificial_2014, bhat_recognition_2016, vyas_automatic_2016, zaidi_automatic_2020, al-qatab_classification_2021, zaidi_deep_2021, sahane_dysarthric_2021, sahu_analysis_2022, jothieswari_enhancing_2024, moro-velazquez_forced_2019, godino-llorente_automatic_2004}. LPC features describe the distribution of energy across frequency bands and are directly related to the resonant properties of the vocal tract, making them suitable for analyzing articulatory features like formant frequencies and tongue positioning. In contrast, MFCCs and PLP features are obtained by transforming the spectral envelope into the cepstral domain, which effectively separates source and filter components while capturing smoothed representations of the vocal tract shape. When extended with temporal dynamics, these features can track transitions and instability in articulation patterns. To enhance robustness against channel variability and noise, the Relative Spectral Transform PLP features have also been considered in~\cite{mfcc1, mfcc2}. Furthermore, the fusion of modulation spectra features or glottal features with MFCCs has additionally been explored to further improve performance~\cite{mfcc4, kadiri2019analysis}.

\paragraph*{Spectro-temporal sparsity} Since pathological speech can be breathy, semi-whispery, and is 
characterized by abnormal pauses and imprecise articulation, it can be expected that its spectro-temporal sparsity differs
from the spectro-temporal sparsity of neurotypical speech.
To characterize spectro-temporal sparsity, various sparsity-based features have been introduced in \cite{kodrasi_spectro-temporal_2020, kodrasi_statistical_2018}.
Although such features have been shown to be discriminative of various speech disorders\cite{kodrasi_spectro-temporal_2020, kodrasi_statistical_2018, speech_mode_impact, descrimination_ina}, they are highly sensitive to environmental artefacts such as noise and reverberation.

\subsection{Time-Frequency Representations} 
Input representations such as the STFT and its variants allow for the analysis of speech signals in both time and frequency domains, providing valuable insights into speech characteristics such as pitch, formant shifts, and spectral irregularities. The STFT is often employed due to its ability to capture dynamic changes in the signal over time, which is critical for identifying variations in speech patterns linked to speech disorders~\cite{parvanehthesis, bhat_dysarthric_2018, rathod_whisper_2023, harvill_synthesis_2021, naeini_improving_2024, vattis_sensitive_2024, yue_acoustic_2022, zhao_personalizing_2021, wu_sequential_2021, yue_raw_2022, soleymanpour_synthesizing_2022, geng_speaker_2022, vora_hybrid_2022}. Furthermore, wavelet transforms and the Continuous Wavelet Transform offer better time-frequency localization, which is especially beneficial for analyzing transient and non-stationary features of pathological speech \cite{vasquez2017convolutional}.

\subsection{Raw Waveform Representation}
While handcrafted acoustic features and time-frequency representations have demonstrated strong performance, researchers have also explored end-to-end pathological speech detection using raw input representations, eliminating the need for handcrafted features or time-frequency representations. Studies such as \cite{rawspeech1, rios-urrego_end--end_2022, narendra_raw} have shown that raw waveform-based methods can capture discriminative pathological patterns and outperform traditional feature-based approaches. However, these methods may require more training data compared to other approaches for a robust performance.

\subsection{Self Supervised Embeddings}
Even though handcrafted acoustic features, time-frequency, and raw waveform input representations have achieved promising results in the analysis of pathological speech, their performance remains limited. \textcolor{black}{The current state-of-the-art input representations for pathological speech primarily consist of latent embeddings derived from self-supervised models such as wav2vec2 \cite{baevski2020wav2vec}, HuBERT~\cite{hubert}, or WaveLM~\cite{waveLM}.
Research in e.g.,~\cite{jiang_perceiver-prompt_2024, kadiri_investigation_2024, tirronen_hierarchical_2023, kheddar_deep_2023, fritsch_novel_2023, ribas_automatic_2023, gengfly_2022, mujtaba_inclusive_2024, soky_domain_2023, maisonneuve_towards_2024, weise_impact_2023, liu_automatic_2024, kheirkhahzadeh_speech_2023, javanmardi_wav2vec-based_2023, baskar_speaker_2022, cullen_improving_2022, nguyen_exploring_2024, lin_cfdrn_2023, speech_mode_impact, wang_benefits_2023, hu_exploring_2023, wang_enhancing_2024, 
 hernandez_cross-lingual_2022, spijkerman_using_2022, shor_trillsson_2022, favaro_interpretable_2023} has shown that these advanced models are effective in capturing complex speech patterns and nuances, leading to improved performance in tasks like detection and recognition.}
However, ongoing research continues to refine these embeddings and assess their robustness and generalizability across diverse pathological speech conditions, tasks, and applications.

 \section{Automatic Pathological Speech Detection}
 \label{detection}
As discussed in \Cref{intro}, clinicians traditionally rely on laborious and time-consuming auditory-perceptual measures to diagnose speech impairments accurately. To address the various challenges associated with auditory-perceptual assessments, there has
been a growing interest in the research community to develop automatic approaches for diagnosing pathological speech.
The goal of these approaches is to enhance the detection accuracy to match or surpass human accuracy, thereby ensuring more consistent and reliable diagnosis. 
Automated systems analyze speech patterns and identify anomalies indicative of these neurodegenerative conditions by leveraging classical machine learning and deep learning models with input representations such as the ones discussed in \Cref{feats}. 
In the following, we briefly summarize various classical machine learning-based and deep learning-based approaches. 

\subsection{Classical Machine Learning-based Approaches}
In recent years, numerous studies have investigated the use of classical machine learning classifiers combined with hand-crafted acoustic features for detecting pathological speech, yielding promising results. For instance, \cite{tsanas_novel_2012} and \cite{vaiciukynas_detecting_2017} employed Random Forests and Support Vector Machines (SVMs) on a range of dysphonia measures, achieving strong classification performance on relatively small speaker datasets. Similarly, \cite{jothilakshmi_automatic_2014} applied Gaussian Mixture Models and Hidden Markov Models to cepstral features for detecting dysarthric speech in an Indian Tamil language dataset. \citet{orozcoarroyave_spectral_2015} further demonstrated that cepstral coefficients are effective and robust when used with SVMs for detecting dysarthric speech in a Spanish dataset. 
While these findings are encouraging, their scalability to larger, more diverse datasets remains uncertain due to the variability of speech pathologies across populations. Additionally, despite the effectiveness of various spectral and cepstral features, the field still lacks consensus on which features are the most discriminative and generalizable across different conditions and datasets.

Recognizing the importance of developing discriminative feature representations, the research community has devoted significant efforts to this area. For example, \cite{novotny_automatic_2014} investigated the use of articulatory features for pathological speech detection and achieved promising results on a small dataset of 24 Czech speakers. However, the limited sample size restricts the generalizability of the findings, and extracting articulatory features at scale remains technically challenging. 
In a more comprehensive effort, \cite{wang_towards_2016} utilized an extensive feature set including $6,373$ acoustic, $3,600$ articulatory, and $4$ sensory features to detect ALS, reporting state-of-the-art performance on a dataset of $22$ speakers. \citet{norel_detection_2018} employed SVMs with openSMILE features to detect pathological speech in a larger cohort of 123 Hebrew speakers.  \citet{prabhakera_dysarthric_2018} found that incorporating glottal features alongside openSMILE features improved dysarthric speech detection performance.
Further advancements were made in \cite{kodrasi_spectro-temporal_2020} by introducing spectro-temporal sparsity features, which outperformed temporal sparsity features in classifying dysarthric speech using SVMs. Additionally, \cite{janbakhshi_subspace-based_2021} applied Grassmann discriminant analysis to spectro-temporal subspaces, leveraging singular value decomposition informed by clinical insights into pathological distortions.

A common limitation across these studies is their reliance on relatively small speaker sets and their susceptibility to biases related to sex, age, recording conditions, or language. To evaluate generalization across different domains, \cite{gillespie_cross-database_2017} conducted a cross-database study, demonstrating that an SVM trained on one dataset suffered a significant performance drop when evaluated on another dataset. This highlights the critical need for models that are robust to distributional shifts, including those stemming from differences in dataset characteristics and recording environments.
Moreover, classical machine learning approaches heavily depend on hand-crafted features, which may not fully capture the nuanced and abstract cues associated with pathological speech. These methods also tend to overlook important metadata such as speaker identity, sex, and language.

\subsection{Deep Learning-based Approaches}

Several studies have explored CNN-based architectures for pathological speech detection, leveraging their ability to extract local patterns from time-frequency representations. For instance, \cite{chandrashekar_spectro-temporal_2020, mani_sekhar_dysarthric-speech_2022} applied CNNs to dysarthric speech, achieving promising results on small datasets of control speakers and ALS patients. However, these models exhibited significant inter-speaker variability in performance, suggesting a lack of robustness across individuals. This limitation is particularly critical in clinical contexts where model reliability must generalize across diverse patient populations.
Building on CNN-based models, \cite{janbakhshi_experimental_2022} incorporated phase-based features, i.e., the modified group delay and instantaneous frequency spectra, as complementary inputs to magnitude spectra. This multi-representational approach improved the CNN performance for pathological speech detection. Similarly, \cite{kodrasi_automatic_2021} introduced temporal envelope features, reinforcing that temporal dynamics encode critical pathological cues. {\textcolor{black}{However, these representation augmentations increase model complexity and demand careful preprocessing, which may hinder real-world deployment.}}
While most approaches focus solely on binary classification of pathological versus neurotypical speech, multi-task learning has shown potential for capturing broader articulatory patterns. \citet{vasquez_correa_multitask_2018} introduced a multi-task system based on CNNs that simultaneously learned to classify pathological speech and predict $11$ articulatory deficit attributes. This approach yielded improved generalization across speakers by constraining the learned representations with related speech tasks, demonstrating the utility of auxiliary objectives in enhancing model robustness.

While CNNs excel at spatial pattern recognition, they lack mechanisms to capture temporal dependencies inherent in speech. Recurrent architectures, particularly LSTMs, have been employed to address this gap \cite{mayle2019diagnosinglstm, millet_learning_2019, bhat_automatic_2020, rios-urrego_end--end_2022}. These models demonstrate improved performance but often require larger datasets to train effectively. Moreover, although the LSTM methodology in \cite{mayle2019diagnosinglstm} showed promising results, the model has been evaluated only on syllables and its generalization to other speech modes remains unexplored.

More recently, there has been a shift toward end-to-end models using self-supervised learning (SSL) to bypass manual feature engineering. SSL models like wav2vec 2.0 leverage raw waveform inputs to learn contextual embeddings, achieving state-of-the-art results across several pathological speech datasets~\cite{javanmardi_pre-trained_2024, javan_mardi_fine_tuneing, janbakhshi_ua}. These models outperform classical approaches in both accuracy and scalability. However, despite their empirical success, SSL-based models function as black boxes, raising concerns about clinical interpretability and decision transparency.

\section{Pathological Speech Recognition}
\label{pasr}
ASR systems have significantly advanced over the years, demonstrating impressive results in converting raw speech signals into their corresponding textual form. This has resulted in the widespread use of various interactive devices, including smartphones and voice assistants. However, they often fail to recognize low resource pathological speech, including speech from patients suffering from various neurodegenerative conditions such as PD or ALS~\cite{de2019impact}. 
\par 
To alleviate this problem, several attempts have been made in improving the performance of ASR systems for pathological speakers. 
For example, \cite{takashima_two-step_2020} proposed a two-step speaker adaptation method. In the first step, a model trained on extensive control speech data is fine-tuned for dysarthric ASR. In the second step, this fine-tuned model undergoes further adaptation to a specific dysarthric speaker. Additionally, \cite{takashima_dysarthric_2020} proposed the additive angular margin loss to address intra-class variation among dysarthric speakers, demonstrating promising results on Japanese speakers. 
~\citet{green_automatic_2021} also showed that personalized ASR systems fine-tuned on pathological speech exhibit better recognition performance compared to speaker-independent ASR systems. However, the reliance on specific speakers in personalized ASR systems poses challenges for generalization settings.~\citet{hermann_dysarthric_2020} utilize lattice-free maximum mutual information to mitigate insertion errors, which are otherwise prevalent due to the slow speaking rates of individuals with dysarthria. 
In a different approach,~\citet{xiong_source_2020} utilized transfer learning and found that speaker-based data selection leads to negative transfer. They recommended using utterance-based data selection with an entropy distribution to enhance recognition.~\citet{yue_autoencoder_2020} further implemented a multi-task learning (MTL) framework through auto-encoder joint learning, utilizing bottleneck features on out-of-domain data. Employing MTL in pathological ASR showed a lower word error rate compared to its single-task counterpart.
~\citet{shahamiri_speech_2021} demonstrated that state-of-the-art ASR systems for pathological speech are significantly impacted by phoneme inaccuracies. To address this, they proposed Speech Vision, a transfer learning paradigm that converts word utterances into visual feature representations, aiming to recognize the shape of the word rather than relying on phonemes. 

A comprehensive overview of ASR systems for pathological speakers in terms of progress and challenges is provided in~\cite{liu_recent_2021}, where it is shown that pathological speech recognition performance can be significantly improved through a combination of neural architecture search, data augmentation, speaker adaptation, and multi-modal learning. These techniques address challenges such as limited training data, high inter-speaker variability, and reduced speech intelligibility. 
It should be noted that due to the large number of parameters in state-of-the-art ASR models, fine-tuning them on low-resource pathological speech data can be highly expensive. However, leveraging fine-tuning techniques such as Low-Rank Adaptation (LoRA) offers a more efficient alternative, reducing computational costs and making training more feasible~\cite{hu_lora_2021, song_lora-whisper_2024}. Additionally, LoRA's modular approach might make it easier to adapt the fine-tuned models to new speakers, providing flexibility in real-world applications or clinical scenarios.

\section{\textcolor{black}{Intelligibility Enhancement \\ of Pathological Speech}}
\label{enhacement}

Pathological speech enhancement refers to improving the intelligibility and quality of speech affected by impairments such as dysarthria. The benefits of such enhancement are twofold. First, enhanced speech can ease human-human and human-machine communication for pathological speakers, promoting their social and digital inclusion. Second, enhancement methods can be leveraged for data augmentation (cf. Section~\ref{dataaug}), aiding in the development of robust models for pathological speech processing.

Early efforts in pathological speech enhancement focused on explicitly correcting articulatory and acoustic deficiencies. For instance, \cite{kain04_ssw} improved intelligibility by restructuring formant trajectories to better match intended speech targets. \citet{rudzicz_adjusting_2013} applied a range of techniques including pronunciation correction, phoneme insertion, tempo adjustment, and removal of disfluencies. \citet{hosom_intelligibility_2003} modified short-term spectral features to enhance word-level intelligibility, though their approach was speaker-dependent. \citet{lalitha_kepstrum_2010} proposed a speaker-specific Kepstrum-based method, while \cite{dhanalakshmi_intelligibility_2015} introduced a two-stage framework that combined ASR with speech synthesis to correct mispronunciations.

Several mapping-based approaches aiming to transform dysarthric speech into more intelligible forms have also been explored. 
For example, \cite{wang_dysarthric_2022} proposed a speech enhancement method where a CNN is trained to directly map dysarthric utterances to their control counterparts. At the feature level, techniques such as LPC mapping and frequency warping of LPC poles have been explored in \cite{kumar_improving_2016} and \cite{roy_towards_2019}. Additionally, \cite{bhat_dysarthric_2018} investigated feature-level mapping using time-delay neural networks.

More recent advances leverage voice conversion models for pathological intelligibility enhancement. For instance, \cite{yang_improving_2020} framed the task as a style transfer problem and employed GANs to convert dysarthric to typical speech. Similarly, \cite{halpern_objective_2021} utilized CycleGAN for dysarthric-to-typical speech conversion, and \cite{purohit_intelligibility_2020} applied DiscoGAN to map pathological and typical speech features at the acoustic level. Among these, \cite{prananta_effectiveness_2022} showed that time-stretching combined with MaskCycleGAN outperformed other GAN-based models in intelligibility enhancement, although the technique’s practicality remains limited due to computational complexity. In a more comprehensive system, \cite{matsubara_high-intelligibility_2021} integrated Transformer-TTS, CycleVAE-VC, and LPCNet to generate highly intelligible dysarthric speech, albeit at the expense of naturalness. Additionally, \cite{wang_end--end_2020} proposed an end-to-end voice conversion system using knowledge distillation for enhanced dysarthric speech synthesis.
To avoid adversarial training instability in GAN-based voice conversion models, neural encoder-decoder architectures have also been investigated for improving pathological speech intelligibility. For instance, \cite{wang_unit-dsr_2024} introduced Unit-DSR, which converts dysarthric speech into discrete linguistic units and then reconstructs speech from these normalized units using a neural vocoder.

In summary, the evolution of pathological speech enhancement methods reflects a progression from early signal processing and mapping approaches to advanced generative models and neural architectures. While GAN-based models have significantly improved intelligibility, challenges remain in balancing enhancement quality, naturalness, and computational efficiency. Neural encoder-decoder frameworks show promise in addressing these challenges, marking an important direction for future research.

\section{Intelligibility and Severity Assessment \\of Pathological Speech}
\label{intelli_assessment}

{\emph{Intelligibility.}} \enspace
Intelligibility of pathological speech is a critical indicator for evaluating the effectiveness of speech therapy and tracking the progression of various disorders. 
To reduce the burden of evaluating pathological speech intelligibility in clinical practice, automatic approaches have been proposed in the literature.
Automatic pathological speech intelligibility assessment methods are typically categorized into two main approaches, i.e., blind and non-blind approaches~\cite{parvanehthesis}. In blind approaches, the objective is to assess the intelligibility of impaired speech without exploiting reference neurotypical speech data~\cite{paja_automated_2012, martinez2013dysarthria, kim2014speech, hummel2011spectral, falk2012characterization, haderlein, fletcher2017predicting}. These approaches primarily focus on extracting acoustic features such as jitter, fundamental frequency, shimmer, formant frequencies, etc., that are believed to be closely correlated with speech intelligibility. These features are then used in regression models to estimate the intelligibility of pathological speech. Non-blind approaches, by contrast, rely on intelligible speech from neurotypical speakers as a basis for estimating the intelligibility of pathological speech~\cite{haderlein2004automatic, middag2008objective, middag2009automated, middag2010towards, maier2009peaks, nuffelen2009speech, bocklet2012automatic, martinez2015intelligibility, imed2017automatic, kalita2018intelligibility}.
Such approaches typically use features extracted from ASR systems, which have been trained on large amounts of control speech, to train regression models to estimate the intelligibility of pathological speech.
 To avoid the burden of collecting and transcribing a large amount of neurotypical speech data required for such systems,~\cite{janbakhshi2019pathological} proposed the pathological short-time objective intelligibility measure (P-STOI) adapted from the speech enhancement domain. The P-STOI measure first calculates an utterance-dependent fully intelligible representation from a small set of control speakers.
 The intelligibility of the pathological utterance is then evaluated by quantifying its divergence from this reference representation in terms of the short-time spectral correlation.
 While advantageous, P-STOI requires recordings of the same utterance from intelligible control speakers, which may not always be available. To mitigate this issue, \cite{janbakhshi2020synthetic} developed a method to generate synthetic reference speech for assessing pathological speech intelligibility. 
In a different approach, \cite{janbakhshi2020automatic} introduced sub-space based intelligibility measures based on the premise that dominant spectral patterns in pathological speech deviate significantly from those of intelligible speech. 
Although such measures result in a lower performance than measures exploiting neurotypical intelligible speech, they can be directly used in practical scenarios where such speech material in not available or easy to generate.

{\emph{Severity.}} \enspace
Besides intelligibility assessment, severity assessment is another important research area where developing tools for this purpose could greatly assist in automatizing the tedious process of screening patients and categorizing them into different subgroups based on the severity level. The methods developed in the literature for this purpose can be categorized into two categories, i.e., traditional machine learning-based approaches using the Mahalanobis distance classifier~\cite{paja_automated_2012}, SVMs~\cite{yeo_automatic_2021}, GMMs~\cite{kadi_fully_2016}, or decision trees~\cite{dubbioso_precision_2024}, and deep learning-based approaches~\cite{soleymanpour_increasing_2021, joshy_automated_2021, gupta_residual_2021, joshy_automated_2022, joshy_dysarthria_2023-1, radha_variable_2024, sajiha_automatic_2024}. 
While deep learning approaches aim to automatically extract acoustic cues correlated with severity from raw or minimally processed speech signals, traditional machine learning approaches for severity assessment rely on (clinically informed) handcrafted acoustic features.
For example, motivated by auditory processing knowledge, \cite{gurugubelli_perceptually_2019} introduced perceptually enhanced single frequency cepstral coefficients for assessing the severity of pathological speech.
\citet{vasquez-correa_towards_2018} showed that articulation features extracted from continuous speech signals to create i-vectors were advantageous in quantifing the dysarthria severity level.
Based on the knowledge that pathological speakers often exhibit irregular rhythm patterns in their speech, \cite{hernandez_dysarthria_2020, hernandez_prosody-based_2020} explored rhythm-based features for severity assessment.

Besides speech impairments, patients with various pathological conditions often display distinct facial expressions. To harness these visual features, \cite{tong_automatic_2020} introduced the first audio-visual pathological severity classification system using CNNs.
To leverage metadata information such as age, sex, and type of pathological condition, \cite{joshy_dysarthria_2023} proposed a multi-head attention-based MTL framework. This approach jointly optimizes severity, type, sex, and age classifications for pathological speech, thereby enhancing the robustness of latent features across these additional factors.
Recently, there has been growing interest in using SSL embeddings to measure pathological speech severity. This approach is promising due to the scarcity of labeled pathological data and the ability to leverage unlabeled data and metadata from other datasets, making it well-suited for resource-constrained settings~\cite{rathod_whisper_2023, javanmardi_wav2vec-based_2023}.

\section{Data Augmentation for \\ Pathological Speech Applications}
\label{dataaug}
To mitigate overfitting in deep learning models dealing with the low resource pathological speech data, data augmentation techniques have been used for various tasks. Existing approaches to data augmentation in pathological speech are broadly based on traditional strategies (such as incorporating noise, reverberation, or multiple datasets), perturbation strategies, voice conversion, and text-to-speech (TTS) synthesis. 

\citet{takashima_end--end_2019} employed a strategy that involved combining diverse pathological speech data from multiple languages to increase the number of data samples. Their results indicate that such a traditional data augmentation approach can be advantageous for pathological ASR. 
Based on the temporal and speed differences between pathological and control speech, \cite{vachhani_data_2018, geng_investigation_2022, yue_raw_2022} have explored vocal tract length perturbation, tempo perturbation, and speed perturbation as data augmentation approaches for pathological ASR. 
Similarly, \cite{xiong_phonetic_2019} introduced a data augmentation technique that adjusts the phonetic-level tempo of healthy speech to resemble atypical speech, and vice versa. Their findings demonstrated that the former approach is more effective for pathological ASR. 
More recently, voice conversion and TTS systems are also commonly employed to generate pathological speech from healthy speech \cite{yang_improving_2020, leung_training_2024, hermann_few-shot_2023}. These synthetic samples serve a dual purpose, i.e., they can be used for speech enhancement, improving the clarity and quality of pathological speech; and they can be used for expanding the dataset, thereby enhancing the diversity of training data. Increased data diversity is beneficial to avoid overfitting and improve the performance of deep learning models in tasks such as pathological speech detection and pathological ASR~\cite{yang_improving_2020, leung_training_2024, jin_adversarial_2021, jin_personalized_2024}.
Since pathological speech datasets typically have a limited vocabulary, using a voice  conversion or TTS model can also be used to expand the set of out-of-vocabulary words~\cite{harvill_synthesis_2021}.
Recently,~\cite{wang_enhancing_2024} investigated different data augmentation strategies in pathological ASR, demonstrating that GAN-based conversion methods are more effective than perturbation-based augmentation approaches. However, a comprehensive investigation of the advantages of all data augmentation strategies in various pathological speech processing tasks using multiple datasets is still lacking. 

\section{Challenges and Research Directions in Pathological Speech Processing}
\label{challenges}
\bl{
In clinical settings, the integration of automated speech processing systems for pathological speech analysis is crucial for advancing the diagnosis, therapy, and monitoring of various disorders. While standardized clinical scales are invaluable in perceptually evaluating pathological speech, the incorporation of automated systems can offer additional benefits in terms of objectivity, efficiency, and real-time feedback. Tools such as VoxTester \cite{voxtester}, for instance, provide clinicians with quantifiable speech metrics, including articulation precision and speech rate, which help monitor disease progression and assess the effectiveness of interventions. However, to fully realize the potential of these tools, further research is needed to ensure that they can seamlessly integrate into clinical workflows, offering clinicians user-friendly, reliable, and interpretable outputs. Addressing issues such as clinician training, data security, patient consent, and adaptability to various clinical settings is crucial for the real-world adoption of these systems. Moreover, creating systems that can operate longitudinally, i.e., tracking speech performance over time to capture subtle changes in speech function, would significantly enhance therapeutic outcomes. We believe that addressing the challenges and research directions outlined in the remainder of this section will be key to enabling the seamless integration of these technologies into clinical practice in the future.
}

\subsection{Impact of Speech Mode}
The large majority of previously reviewed deep learning-based pathological speech approaches have been proposed and validated on controlled speech tasks. Controlled speech, also known as non-spontaneous speech, involves utterances produced within a structured context, typically requiring participants to repeat phonetically balanced, carefully crafted texts. This mode is designed to elicit specific pathological biomarkers by standardizing the motor planning and execution demands. Tasks may include reading aloud or repeating scripted phrases.
In contrast, spontaneous speech consists of unplanned utterances, such as storytelling or casual conversation, which reflect real-world communicative behavior. It places greater demands on cognitive planning, articulation, and natural language generation, and may therefore reveal more authentic or varied pathological cues.
While this distinction has received some attention in the context of detection, like in~\cite{speech_mode_impact}, its implications extend across other subfields such as severity assessment, intelligibility prediction, enhancement, and recognition. For instance, severity assessment models trained only on controlled speech may generalize poorly to real-world settings. Similarly, enhancement models may be tuned to the acoustic patterns of controlled tasks, failing to capture the variability in spontaneous speech. 
Incorporating spontaneous speech more broadly into pathological speech research could enhance practical utility and performance across these subfields. Given its ease of collection and alignment with natural communication, spontaneous speech provides a more suitable and informative basis for training and evaluating models~\cite{speech_mode_impact, sheikh2024graph, kaloga2024multiview}. Future work should consider systematically analyzing the effect of speech mode across multiple application areas to ensure generalizability and real-world effectiveness.

 \subsection{Robustness in Pathological Speech Detection}
 \label{robustnes}
 The methods developed so far for automatic pathological speech detection have typically been designed and tested under specific environmental conditions and for a particular language. Each dataset used for pathological speech detection has its own unique variabilities in terms of both inter-speaker and intra-speaker differences. Additionally, different age groups exhibit distinct speaker attributes, adding complexity to pathological speech analysis. Language is another important factor; separate models have been independently developed for different languages. However, there is a need for a more robust model that is language-agnostic (to a possible degree given that pathological cues might be different in different languages), age-agnostic, accent-agnostic, and sex-agnostic to enhance overall effectiveness~\cite{favaro_interpretable_2023}. Additionally, datasets such as TORGO are contaminated with noise. Researchers have shown that models trained on these datasets tend to learn environmental factors rather than focusing on genuine pathological cues~\cite{janbakhshi_ua}.
The development of pathological speech detection models robust to environmental distortions has been very limited, with only a few small studies addressing this area~\cite{wisler_noise_2016, amiritest, amiri_iwanc, ibarra2023towards}. For example, \cite{amiritest} employed a test-time adaptation method to fine-tune pre-trained models on a validation set augmented with the test noise extracted from the test utterance. This method improves the robustness of state-of-the-art pathological speech detection methods, offering a promising solution to deploying such applications in realistic clinical settings.
 Similarly, 
 \cite{amiri24_interspeech} proposed an approach to resolve the noise disparity between the two groups of speakers in the TORGO database, such that models developed on this database learn pathology-discriminant cues instead of noise-discriminant ones.
 Besides robustness to environmental distortions, adversarial robustness of pathological speech detection models is another important topic and research direction. The impact of acoustically imperceptible adversarial perturbations on deep learning-based pathological speech detection models has been explored in \cite{amiri_iwanc}. Results revealed a high vulnerability of such models to adversarial perturbations, with
adversarial training ineffective in enhancing robustness.

\subsection{Improving the Performance of Automatic Pathological Speech Analysis}
\label{gcn}
While many automatic pathological speech detection methods have shown remarkable performance, the exploitation of common attributes such as pathological cues, speaker characteristics, age, and sex across different speakers remains limited. A promising direction is to approach the pathological speech detection problem as a semi-supervised node graph classification task using graph neural networks (GNNs), as demonstrated in \cite{sheikh2024graph}. This approach could involve constructing an inter-speaker graph based on utterances from various speakers, where the graph’s connectivity would help form speaker clusters based on the presence or absence of pathological cues. In this context, domain knowledge can be easily integrated by establishing edges based on factors such as sex, age, and the severity scale of the patients.
Additionally, \bl{the recent availability of longitudinal multimodal data~\cite{favaro2024unveiling} enables novel applications of GNNs in disease monitoring. By representing the patient history as a temporal graph where nodes capture multimodal features (e.g., vocal biomarkers, facial expressivity) and edges encode their dynamic interactions, GNNs can model disease progression through evolving graph topologies.

An additional area that remains under-explored in current research, potentially due to the lack of large datasets, is the adoption of Bayesian frameworks and generative approaches for pathological speech processing. Bayesian frameworks are particularly relevant in healthcare, as they allow the model to estimate uncertainties in its predictions, which is essential for high-risk settings where incorrect predictions can have serious consequences. These models are ideal for situations where speech data may vary considerably due to speaker differences or environmental factors. By providing probabilistic models, Bayesian methods can account for such variability, making them highly suitable for clinical applications where individual differences are pronounced. 
}

\subsection{Privacy}
\label{pri}

In pathological speech-based applications such as detection models or ASR systems, privacy-preserving solutions are crucial for safeguarding sensitive patient data. 
Since these systems process speech that reveals sensitive medical information, ensuring data security and confidentiality is of paramount importance. 
By employing privacy-preserving techniques such as federated learning, differential privacy, and encryption, providers can ensure that individuals' speech data is anonymized and never exposed to unauthorized parties. This not only fosters trust among patients but also complies with stringent data protection regulations, thereby mitigating the risk of breaches. Developing privacy-preserving pathological speech processing systems where one must balance two conflicting goals, i.e., increasing the utility of the models while preserving the privacy of the users, remains an important challenging research directions.

\subsection{Multimodal Pathological Speech Analysis}

Most existing systems primarily focus on leveraging speech cues for detecting pathological speech. However, complementary cues may also exist in visual forms, such as facial expressions or lip movements, which can provide additional insights. \bl{To effectively utilize these visual cues, multimodal self-supervised methods such as AV-HuBERT~\cite{avhubert} could prove advantageous, as they facilitate the integration of multimodal audio and visual information, potentially enhancing the accuracy and robustness of pathological speech detection models. Furthermore, combining other modalities, such as medical \blu{imaging}, health records, brain signals, and textual data, could offer a more complete characterization of disorders}. These multimodal systems may help in capturing underlying neural, visual or linguistic patterns associated with speech pathology, improving pathological speech processing across diverse patient groups. 

\subsection{Explainability and Interpretability}
\label{explain}

In the domain of pathological speech detection, explainability and interpretability are crucial due to their clinical significance. Although these terms are often used interchangeably, it is important to distinguish between them. As defined in \cite{rudin_stop_2019}, interpretable models are designed to be inherently understandable, whereas explainable models provide post-hoc explanations for the decisions made by existing black box systems that are otherwise incomprehensible to humans.

Despite their importance, relatively little attention has been paid to explainability and interpretability in pathological speech detection, with most existing research focusing primarily on improving model accuracy. Nonetheless, a few notable efforts have begun to address this gap~\cite{jiao_interpretable_2017, rowe_acoustic-based_2020, turrisi_interpretable_2022, gimeno2025unveiling, xu_dysarthria_2023}. For example, \cite{jiao_interpretable_2017} mapped high-dimensional acoustic features to binary phonological representations, recognizing that raw acoustic features are difficult to interpret in clinical practice. \citet{xu_dysarthria_2023} applied the SHAP algorithm to identify the most influential features in their model, highlighting the role of consonant-vowel transitions in reversing classification decisions. Similarly, \cite{kaloga2024multiview} used canonical correlation analysis to show that the $0$--$210$ Hz frequency range strongly influences model outputs.

Recently, \cite{favaro_interpretable_2023} found that non-interpretable SSL embeddings outperform interpretable features (e.g., prosodic, linguistic, and cognitive descriptors) in both multilingual and cross-lingual contexts. This highlights a growing challenge, i.e., as models increasingly rely on high-performing but opaque representations like SSL embeddings, there is a pressing need to develop methods that make these models explainable and clinically trustworthy.

\subsection{Large Language Models for Pathological Speech}
\label{llm}
 Given the recent advancements in multimodal large language models (LLMs), which have demonstrated significant progress across a wide range of applications~\cite{hadi_large_2024, wagner_large_2024}, exploring their potential for pathological speech analysis appears to be an inevitable and promising direction. Multimodal LLMs could be leveraged for tasks such as detecting speech impairments, ASR, and enhancing the reconstruction of unintelligible or difficult speech. 
  Their ability to capture complex patterns may lead to more accurate models for personalized therapy, rehabilitation, and assistive communication tools for individuals with speech impairments. 
 Additionally, multimodal LLMs should be explored for their ability to explain their decisions.
 This exploration could bridge the gap between traditional speech processing techniques and state-of-the-art language models, opening up new avenues for more effective and adaptive speech rehabilitation systems. 

\section{Conclusion}
 
This paper provides a comprehensive overview of speech analysis and technologies for pathological speech arising due to neurological disorders, encompassing detection, recognition, intelligibility assessment, and enhancement. Additionally, it compiles a thorough list of both accessible and non-accessible pathological speech datasets, which will serve as valuable resources for future research and accelerate progress in the field. Finally, it outlines potential future research directions, particularly in the context of robust and interpretable models that can be deployed in clinical practice.

\section*{Acknowledgment}

This work was supported by the Swiss National Science
Foundation project CRSII5\_202228 on “Characterisation of
motor speech disorders and processes”.

\bibliographystyle{IEEEtranN}
\footnotesize \small
\bibliography{refs}

\begin{IEEEbiography}[{\includegraphics[width=1in,height=1.2in,clip]{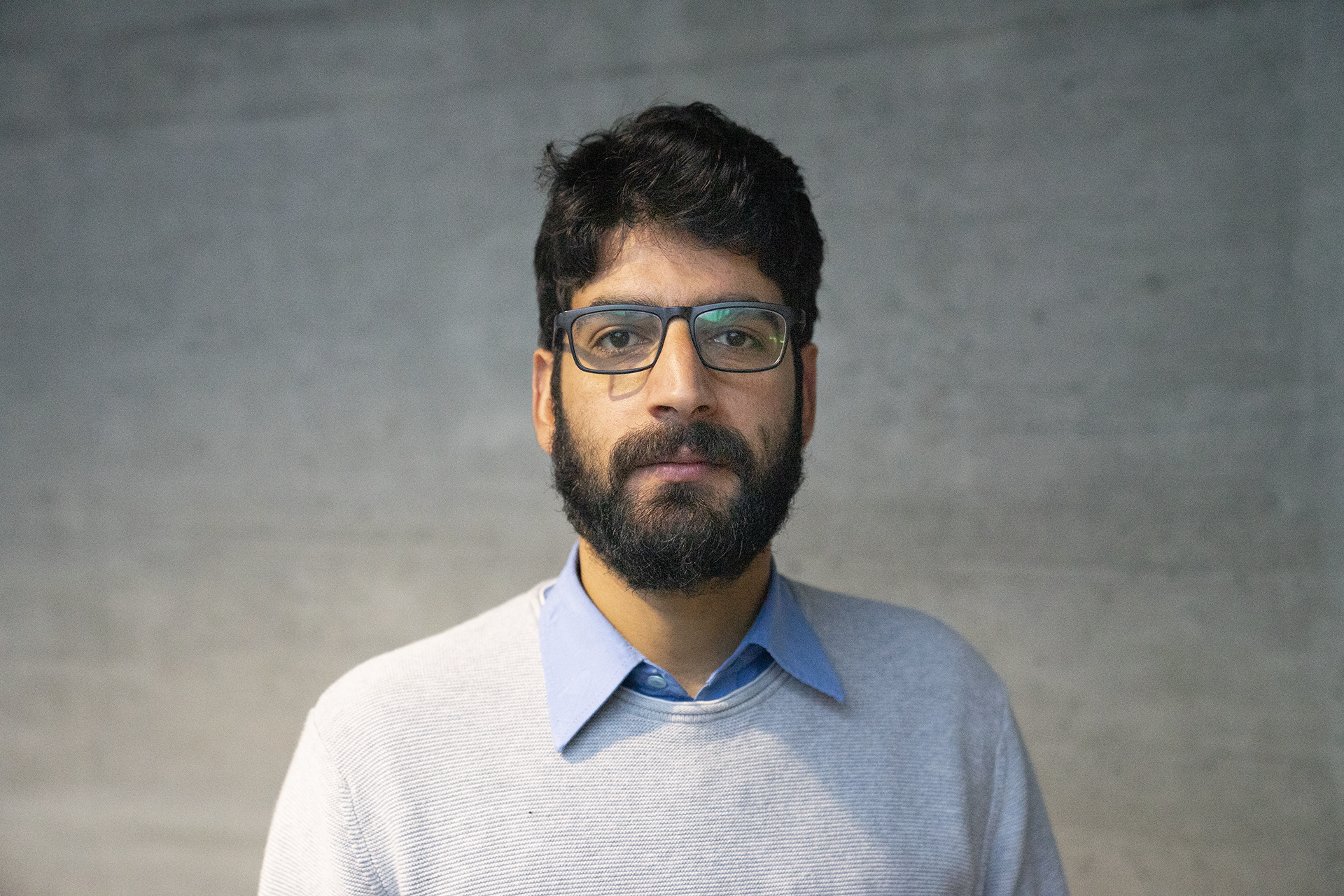}}]
{Shakeel A. Sheikh}
Shakeel A. Sheikh is currently working as a Data Science Innovation Research Scientist in the Oncology Data Science section at Novartis AG, Switzerland. His research focuses on the development and application of advanced data science and machine learning methodologies in oncology. Prior to this, he worked as a postdoctoral research scientist on the ChaSpeePro project at the IDIAP Research Institute, affiliated with EPFL (École Polytechnique Fédérale de Lausanne), Switzerland. He also held a postdoctoral position at the Cluster of Excellence Cognitive Interaction Technology (CITEC), Bielefeld University, Germany, in 2023. He earned his Ph.D. in 2023 from the MULTISPEECH team at LORIA-INRIA, Department of Informatics and Mathematics, Faculty of Sciences, Université de Lorraine, Nancy, France. His doctoral research focused on deep learning techniques for pathological speech. He obtained his M.S. in Informatics from Istanbul University, Turkey, in 2019, and holds a B.Tech in Computer Science and Engineering from the University of Kashmir, Jammu and Kashmir, India in 2015.
\end{IEEEbiography}

\begin{IEEEbiography}[{\includegraphics[width=1in,height=1.25in,clip,keepaspectratio]{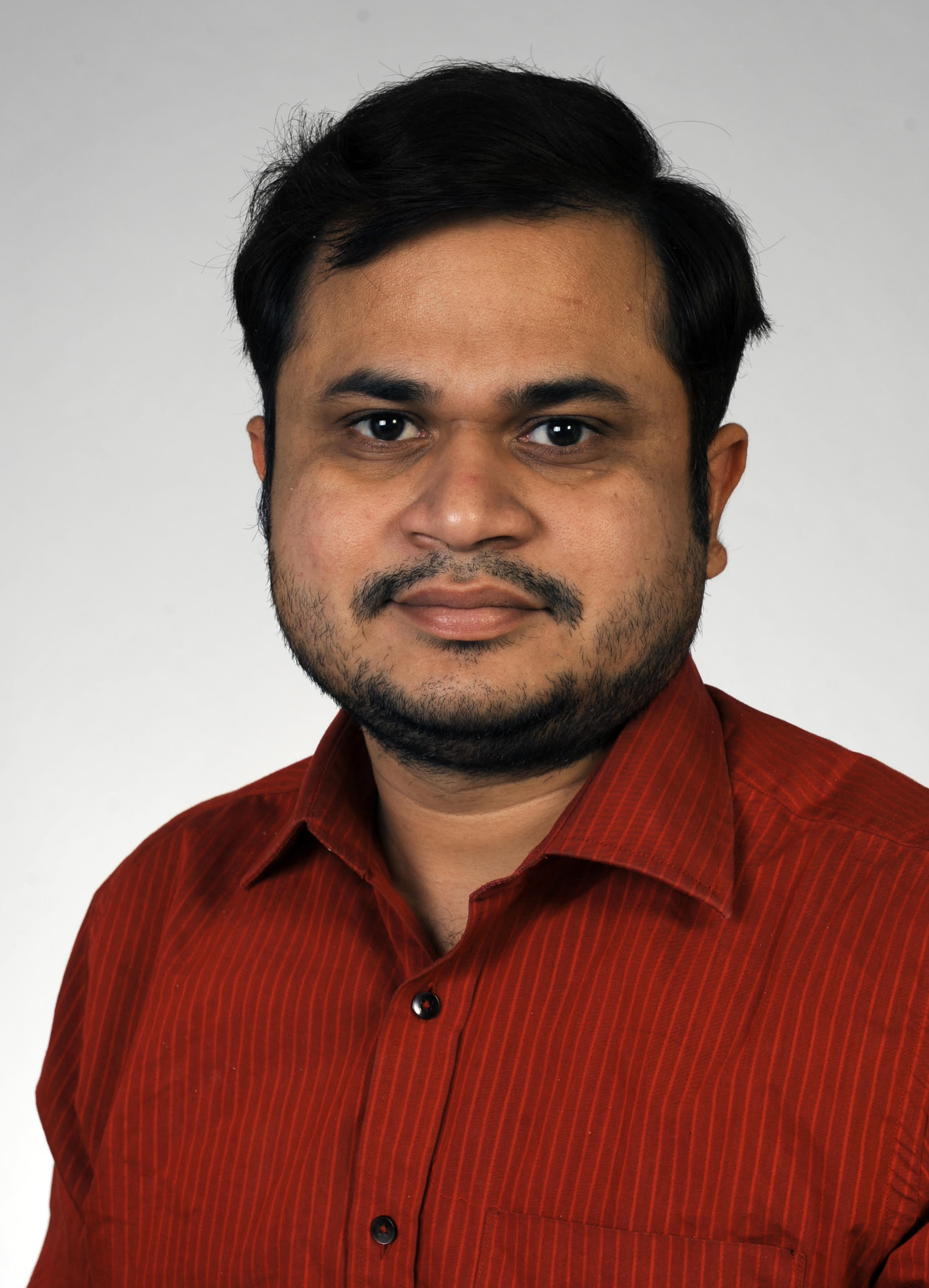}}]{Md Sahidullah} is an Assistant Professor in the Artificial Intelligence and Machine Learning group at the Institute for Advancing Intelligence, TCG CREST. His research interests lie in machine learning and speech/audio processing, with a focus on speech privacy and security, audio analytics for healthcare, and the development of voice-enabled technologies. He has over nine years of post-PhD experience in the field. He received his Ph.D. in Speech Processing from the Department of Electronics and Electrical Communication Engineering at the Indian Institute of Technology Kharagpur in 2015. He holds a B.E. in Electronics and Communication Engineering from Vidyasagar University (2004) and an M.E. in Computer Science and Engineering from the West Bengal University of Technology (2006). He was a postdoctoral researcher at the School of Computing, University of Eastern Finland (2014–2017), and later held a Starting Researcher position with the MULTISPEECH team at Inria Nancy – Grand Est, France (2018–2021). He has contributed to several national and international projects in Finland, France, and the EU. Since 2017, he has co-organized the ASVspoof Challenge, the leading international competition on audio deepfake detection. He has served on technical program committees for top conferences such as ICASSP and INTERSPEECH and is currently a member of the editorial boards of IEEE/ACM Transactions on Audio, Speech and Language Processing, IEEE Journal of Biomedical and Health Informatics, and Computer Speech \& Language.
\end{IEEEbiography}

\begin{IEEEbiography}[{\includegraphics[width=1in,height=1.25in,clip,keepaspectratio]{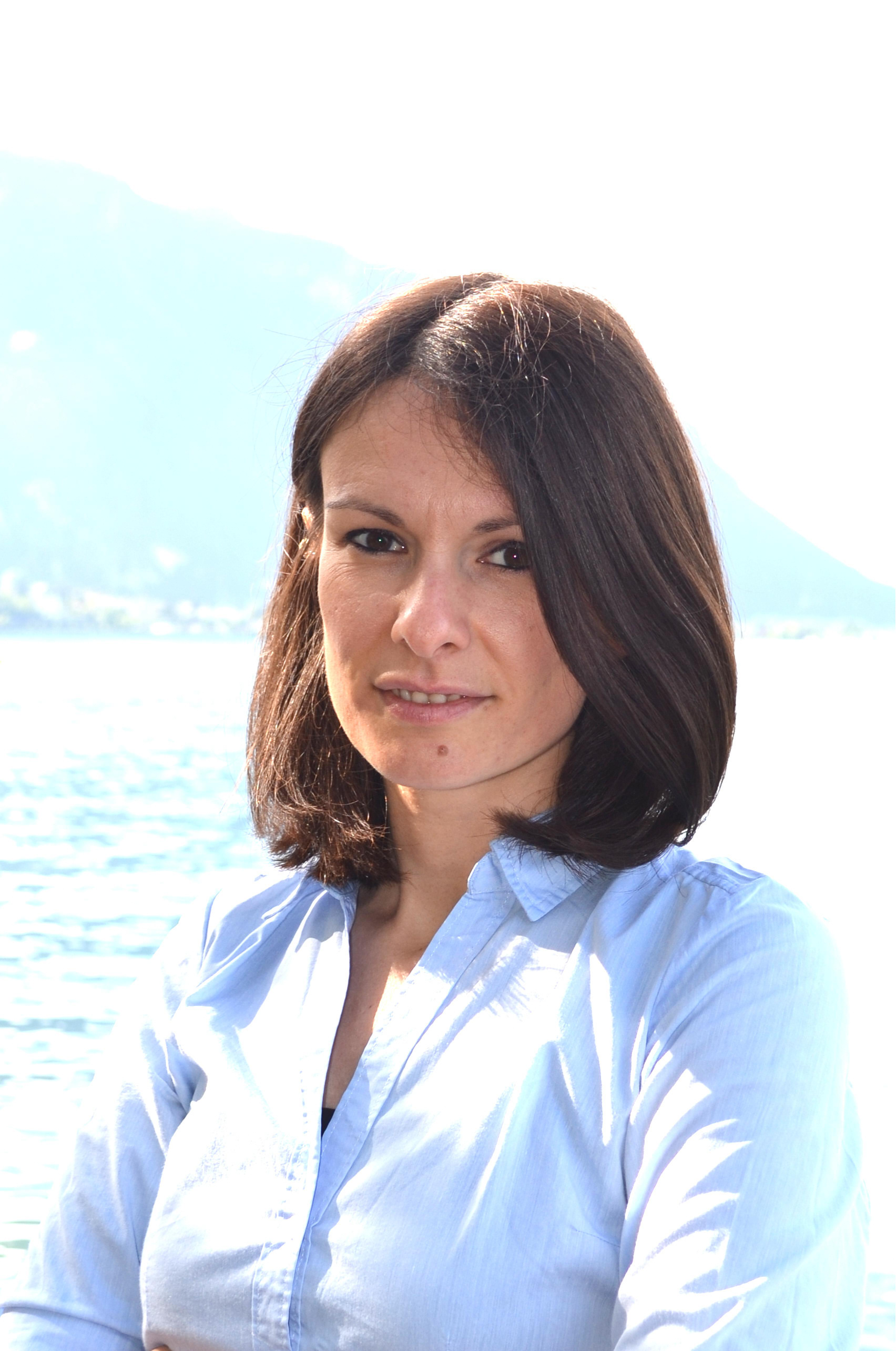}}]{Ina Kodrasi} (Senior Member, IEEE) received the M.Sc. degree in Communications, Systems, and Electronics from Jacobs University Bremen, Germany, in 2010, and the Ph.D. degree from the University of Oldenburg, Germany, in 2015. From 2015 to 2017, she was a Postdoctoral Researcher at the University of Oldenburg, working on multi-microphone speech dereverberation and noise reduction. Since December 2018, she has been with the Idiap Research Institute, Switzerland, where she leads the Signal Processing for Communication Group. Her research interests include signal processing, multichannel processing, pathological speech processing, and machine learning. Dr. Kodrasi received the ITG Best Paper Award in 2019 and the EURASIP Best Ph.D. Award in 2020. She has served as a member of the IEEE Signal Processing Society Technical Committee on Audio and Acoustic Signal Processing and the EURASIP Technical Area Committee on Acoustic, Speech and Music Signal Processing. Since 2023, she has been serving as an Editor for the IEEE/ACM Transactions on Audio, Speech, and Language Processing and the EURASIP Journal on Audio, Speech, and Music Processing.
\end{IEEEbiography}

% \begin{IEEEbiography}[{\includegraphics[width=1in,height=1.25in,clip,keepaspectratio]{fig1.png}}]{IEEE Publications Technology Team}
% In this paragraph you can place your educational, professional background and research and other interests.\end{IEEEbiography}

\end{document}